\documentclass{emulateapj}
\usepackage{graphicx}
\usepackage{natbib} 
\usepackage{amsmath}

\usepackage[ruled, noend, slide]{algorithm2e}

\newcommand{\Pa}{{\rm Pa}}
\newcommand{\data}{{\rm data}}
\newcommand{\Cov}{{\rm cov}}
\newcommand{\Fd}{{\tilde F}}

\begin{document}

\renewcommand{\arraystretch}{1.2}

\title{Automatic classification  of variable stars in catalogs with missing data}

 \author{Karim Pichara \altaffilmark{1}\altaffilmark{,3}\altaffilmark{,4}, 
 Pavlos Protopapas \altaffilmark{2}\altaffilmark{,3}}
\affil{\altaffilmark{1}Computer Science Department, Pontificia Universidad Cat\'olica de Chile, Santiago, Chile}
\affil{\altaffilmark{2}Harvard-Smithsonian Center for Astrophysics, Cambridge, MA, USA}
\affil{\altaffilmark{3}Institute for Applied Computational Science, Harvard University, Cambridge, MA, USA}
\affil{\altaffilmark{4}The Milky Way Millennium Nucleus, Av. Vicu\~{n}a Mackenna 4860, 782-0436 Macul, Santiago, Chile}


\begin{abstract}
  We present an automatic classification method for astronomical catalogs with missing data.
  We use Bayesian networks, a probabilistic graphical model,
  that allows us to perform inference to predict missing values given observed data and dependency relationships between variables.  
  To learn a Bayesian network from incomplete data, we use an iterative algorithm that utilises sampling methods and expectation maximization to estimate
  the distributions and probabilistic dependencies of variables from data with missing values. 
  To test our model we use three catalogs with missing data (SAGE, 2MASS and UBVI) and one complete catalog (MACHO). We examine how classification 
  accuracy changes when information from missing data catalogs is included,  how our method compares to traditional missing data approaches and at what computational cost. 
   Integrating these catalogs with missing data we find that  classification of variable objects improves by few percent and 
   by 15\% for quasar detection while keeping the computational cost the same.
\end{abstract}


\newpage

\section{Introduction}
Classifying objects based on their features (e.g.: color, magnitude or any statistical descriptor) dates back in the 19th century \citep{Rosenberg:1910}. Recently automatic classification methods have become much more sophisticated and necessary due to the exponential growth of astronomical data. In time-domain astronomy, where data is in the form of light-curves, a typical classification method uses features\footnote{we use the term \lq\lq features'' for all the descriptors we may use to represent a light-curve with a numerical vector} of the light-curves and applies sophisticated machine learning to classify objects in a multidimensional features space, provided there are enough examples to learn from (training). After almost a decade since the first appearance of automatic classification methods, many of those methods have produced and continue to produce high fidelity catalogs \citep{Kim:2011, Kim:2012, Bloom:2011, Richards:2011, Bloom:2011, Debosscher:2007, Wachman:2009, Wang:2010}.

To  take full advantage of all  information available, is best to use as many available catalogs as possible. For example, adding u-band or x-ray information while classifying quasars based on their variability is highly likely to improve the overall performance  \citep{Kim:2011, Pichara:2012, Kim:2012}. Because these catalogs are taken with different instruments, bandwidths, locations, times, etc, the intersection of these catalogs is smaller than any single catalog; thus the resulting multi-catalog contains missing values. Traditional classification methods can not deal with the resulting {\em missing} data problem because to train a classification model it is necessary  to have all features for all training members. This can be solved by either selecting the complete intersection of the training members from all catalogs or by deleting the subset of features that are not common to all member. Unfortunately, both methods dramatically reduce the size of the training set because in general most of the features present missing values.

\begin{figure}
  \begin{center}
    \centering
    \includegraphics[width=8cm]{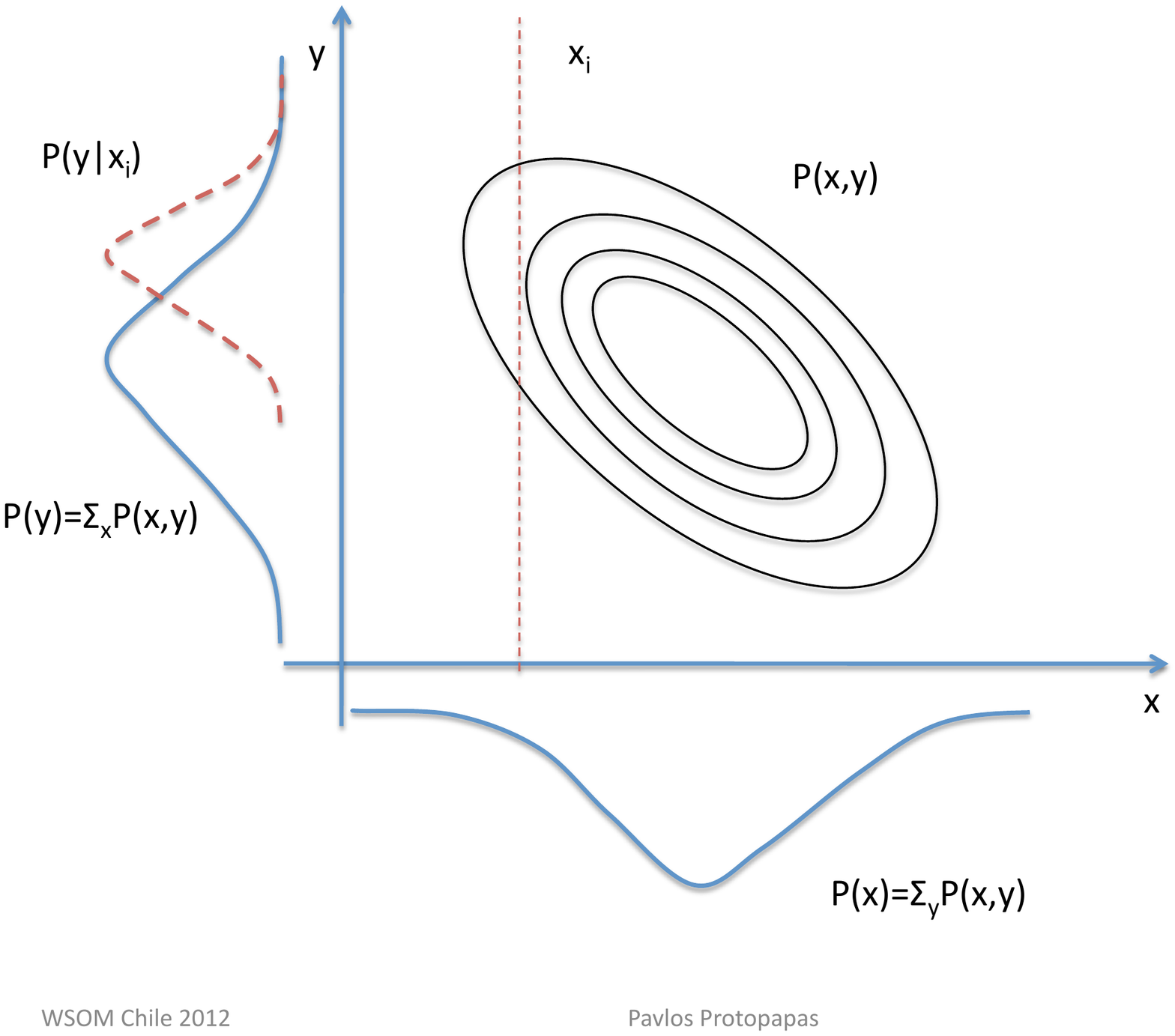}
    \caption{The joint distribution is shown as contours, the marginal distributions are shown in dashed line and conditional distributions in solid lines. Ignoring the joint distribution, one draws from the marginal distribution. However
    knowledge of the joint distribution allows us to draw from the conditional distribution. }
    \label{Fig:TradBN}
  \end{center}
 \end{figure}

Alternatively, one can fill missing data using Monte Carlo approaches where each missing value is drawn from a distribution that is determined from all objects in the training set. However, this approach totally ignores the relationship amongst the features. Fig.~\ref{Fig:TradBN} demonstrates the drawbacks of ignoring such a relationships.  If one draws from the marginal distributions of $x$ and $y$ (shown  with solid blue lines), the fact that $x$ and $y$ are correlated is not taken into account. In principle if we knew that $x$ takes the value $x_i$, then we should be drawing from the conditional distribution, shown with dashed red line.

  One of the main characteristics that an imputation method must have to deal with astronomical catalogs is that the imputation time for new elements with missing data should be very fast, due to the amount of objects in catalogs. There are several data imputation methods presented in the literature (e.g. \citep{Troyanskaya:2001, Stekhoven:2012}). In the method proposed by \citet{Troyanskaya:2001} they used $K$ nearest neighbors to impute the missing data. The basic idea is to select the $K$ nearest neighbors in the space of the non-missing features and then predict the missing variable using a weighted average of the neighbors in that variable. Unfortunately with this method, each time we ask for the imputation of one value we need to find the $K$ nearest objects, which takes a considerable amount of time if we are dealing with millions of object where we want to impute missing data. The method proposed by \citet{Stekhoven:2012} they use one different Random Forest \citep{Breiman2001} model to predict each of the features in the data set. Having $n$ features, they fit $n$ different Random Forests, where the $i$-th Random Forest is trained with the features $\{ 1, \ldots , i-1,i+1,\ldots, n \}$ as predictors and the variable $i$ as a response. To train a Random Forest with variable $i$ as a response, they only use the observed part of variable $i$ in the training set.
  Even though in the paper they propose the iterative model using the whole data (impractical for astronomical catalogs), we might iterate only using a training set and then use the set of forests to impute data in the big catalogs. In astronomical catalogs usually features correspond to astrophysical variables of objects, in many cases, astronomers may want to know an indicator of uncertainty in the prediction, or a probability distribution over the imputation value, unfortunately from Random Forests is hard to directly get uncertainty indicators for continuous responses, given that the model do not provide a distribution over the predictions.

 In this work,  we use (not Naive) Bayesian networks. Bayesian networks are models that represent probabilistic dependency relationships among features using graphs \citep{Pearl:1988}, where nodes represent the features and connections provide information about the probabilistic dependency relationships between features. 
Bayesian networks belong to the family of graphical models and they are very suitable to perform inference on a set of features given 
observations. 
 
  Some recent works in astronomy use  Bayesian network models for automatic classification \citep{Mahabal:2008, Mahabal:2012}. Also \citet{Broos:2011} propose a Machine Learning model to classify  X-ray sources using a naive scheme, where all features are assumed to be independent given the class.

 Usually in catalogs with missing data, the missing features are different depending on the object. That is the main reason to use a model that can deal with evidences that change from  object to object.
 We assume that the nature of missing data is MAR (Missing at Random) or MCAR (Missing Complete at Random). MCAR means that the probability that a feature is missing is independent of the other features
 in observations. MAR means that the missing features may depend on the values of the observed component. MAR and MCAR cases can be handled with our model because we find the dependencies (or independencies)
 between features with BNs.  For NMAR cases (Not Missing at Random), the probability that a feature is missing may depend on the other missing values (for example, no detections when the observed objects are too faint). We do not know any method able to handle NMAR cases without ad hoc distribution for the missing values.  
 
 Note that even in the case we have a complete training set, the resulting classification model will only be able to classify objects that have complete information.
 In other words, we cannot use a classification model to predict objects with missing  features. One option to mitigate this problem is to abstain from predicting for those objects, but that hinders the completeness of the prediction. With the proposed method, we can impute  the data while predicting based on the learned Bayesian network.
		
		
In this work,  we use as a base catalog, the MACHO catalog \citep{Alcock1997ApJa}, and extract 14 features from each lightcurve. We then combine the MACHO catalog with other catalogs containing magnitudes at different wavelengths.
We show how a Bayesian networks in combination with a Random forest classifier can overcome the missing data problem and outperform other methods. Applying this model on a real dataset we are able to generate catalogs of variable source with extremely high fidelity.

Section \ref{sec:bBN} summarizes Bayesian networks, in section \ref{sec:bBNi} we show how BN can be used to infer missing values. Sections \ref{sec:BNi2} and \ref{sec:BNi3} show how we build BN, first with complete data and then with missing values. Section \ref{sec:class}  contains the classification mechanism. Results from experiments with real data are shown in section \ref{sec:Experiments}. Conclusions follow in section \ref{sec:con}.

\section{Theoretical Background}
\subsection{Bayesian Networks}
\label{sec:bBN}
A Bayesian network (BN) is a probabilistic model that belongs into the special class of graphical models. Graphical models deal with uncertain data in presence of latent variables. Latent variables  can include any information that is unobservable, such as the mass of a star. In short, anything that is relevant to explain the observed data but was itself not observed, can become a latent variable. In a graphical model one assumes certain  local statistical 
 dependencies between the random variables that correspond to the latent variables and observed data. 
BNs are directed graphical models, in which the statistical dependency between random variables is based on directional relationships. Another class of graphical modes, not relevant to this paper, are undirected graphical
models, such as Markov random networks.  Many
models that are typically not described as graphical models can be reinterpreted within a
graphical modeling framework. Similarly, many process models or stochastic processes  can be couched as graphical
models.

\begin{figure}
  \begin{center}
   \centering
    \includegraphics[width=8cm]{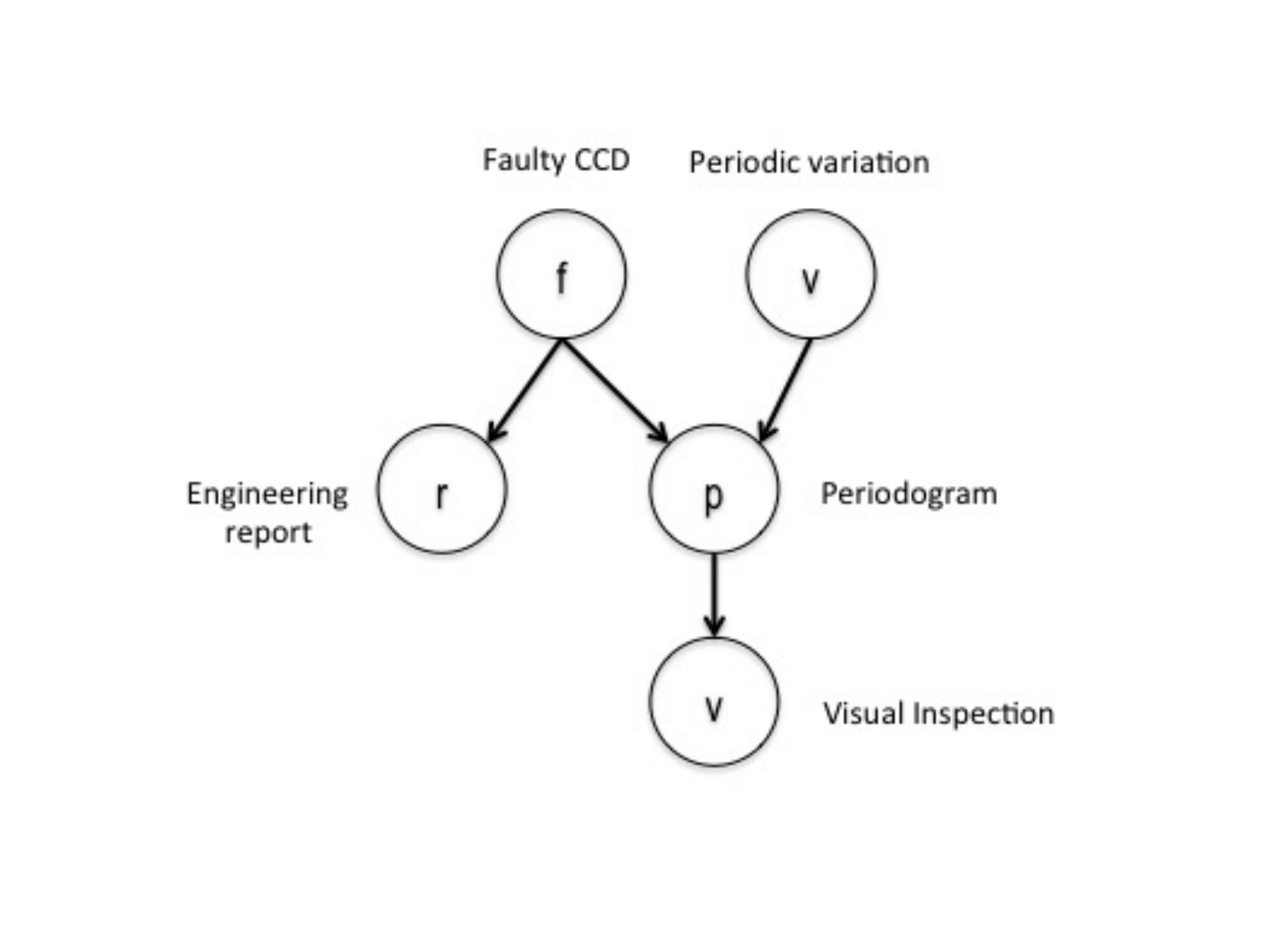}
    \caption{Bayesian network for the periodogram example.}
    \label{fig:BNEx}
  \end{center}
 \end{figure}

To better explain the fundamentals of a Bayesian Network we present a simple example. 
Consider a lightcurve of a source and we examine certain properties 
of the lightcurve and other available information such as
PSF size, magitude, color, etc. In this example we want to determine if the star is periodic or not. 
If the star exhibits periodic behavior that is larger than the error a standard and assume
the source has been observed often and for long time, a simple periodiogram would 
flag this source as periodic very reliably. Unfortunately a faulty CCD or unreliable 
electronics can mimic the periodic behavior which can  fool the periodiogram. Regular engineering
reports can reveal such behavior. Finally, you want to confirm every detection with a 
visual inspection.  
To model this situation, we can use a Bayesian network
as shown in fig. \ref{fig:BNEx}, where nodes are the
random variables and arrows indicate conditional dependencies between variables. 
The network encodes the intuition that the status of the periodiogram results 
depend on faulty CCD or actual periodic variation, and that the final call depends
on the results of the visual inspection that only happens if the periodiogram 
indicates that the source is periodic. It is useful to think of these conditional dependencies as causal relationships
between variables, periodic behavior might cause the periodiogram to flag a light curve as 
periodic, which in turn might pass the visual inspection.  However, you should keep in mind
that Bayesian networks can also be constructed in the absence of any causal interpretation.
This is how \cite{Pearl:94} originally thought of Bayesian network as a way to 
reason probabilistically about causes and effects.

 More formally, let $S = \{ x_{1},\ldots,x_{n} \}$ be a set of data instances (these are the lightcurves), each one described with a set of $D$ features $\{F_1,\ldots,F_D\}$ (these are the lightcurve features and/or brightness magnitudes). Each instance $x_{i}$ is represented as a vector $x_i = \{ F^i_{1} , \ldots , F^i_{D}\}$. BNs can represent the joint probability distribution $P(F_1,\ldots,F_D)$ of dataset $S$ as a product of factors, where each factor is a conditional probability distribution of each node given its parents in the BN:
    
 \begin{eqnarray}
     P(S) & = & \prod_{i=1}^{n} P(x_i)  =  \prod_{i=1}^{n} P(F^i_1,\ldots,F^i_D)  \nonumber  \\              
              &= & \prod_{i=1}^{n} \prod_{j=1}^{D} P(F^i_j |  {\Pa}^i_{BN}(F_j)  )
\label{Eq:BN_Fact}              
 \end{eqnarray}  

     where ${\Pa}_{BN}(F_j)$ represents the set of parents of variable $F_j$ in the BN and ${\Pa}^i_{BN}(F_j)$ indicate that parents of feature $F_j$ are instantiated in the values of $x_i$.  
 
  One of the main advantages of the BN factorization is that each of the factors involves a smaller number of features, where it is easier to estimate.

\begin{figure}
  \begin{center}
    \centering
    \includegraphics[width=6cm,height=5.5cm]{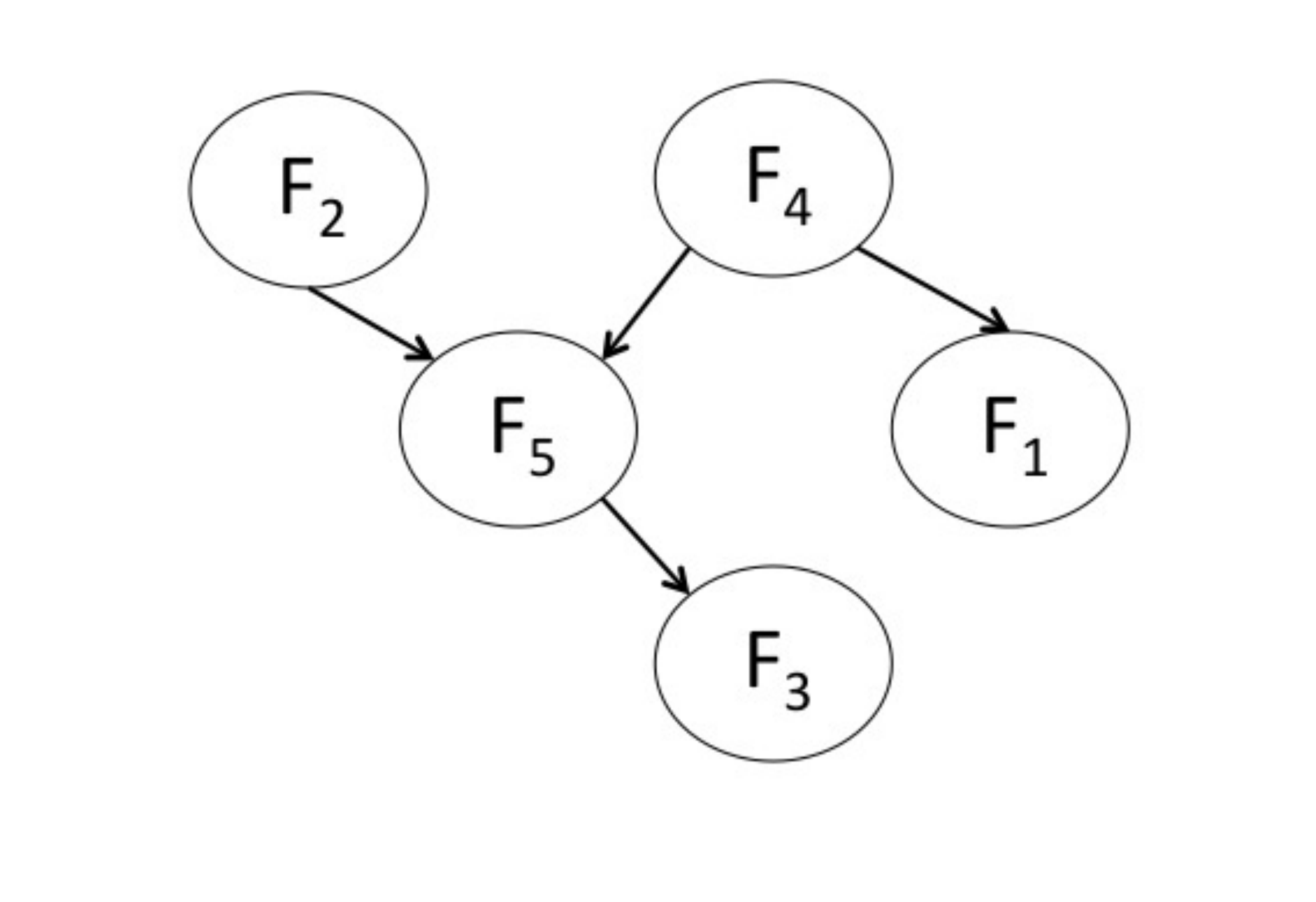}
    \caption{Example of a Bayesian network with features $\{ F_1, \ldots ,F_5 \}$. The joint distribution can be factorized as the product of five probabilities, each one corresponding to the probability of the respective node variable given its parents in the network}
    \label{Fig:BN_example}
  \end{center}
 \end{figure}

For example, in figure \ref{Fig:BN_example} we show a BN in a domain of five features $\{F_1,\ldots ,F_5 \}$. The joint probability distribution can be factorized according to the BN as:  

  \begin{eqnarray}
     P(F_1,\ldots,F_5) &=& P(F_1|F_4)P(F_2)P(F_3|F_5) \cdot \nonumber \\        
                                     & &   \cdot P(F_4)P(F_5|F_2,F_4) \nonumber
   \end{eqnarray}
   
 In this case, instead of estimating a probability distribution over the five dimensional space $(F_1,\ldots,F_5)$ we only need to   estimate simpler distributions, such as $P(F_1|F_4),  P(F_2), P(F_3|F_5), P(F_4)$, and $P(F_5|F_2,F_4)$. 
 
\subsection{Inference in Bayesian Networks}
 \label{sec:bBNi}
BNs are useful to make inference on any unobserved variable given a set of evidence. In our case, we aim to use BNs to predict values of missing features given the observed ones. 

For example, consider the same BN as in Figure~\ref{Fig:BN_example} and suppose we found an object with missing values $F_5$ and $F_2$ (we can observe $F_1,F_3,F_4$). If we want to estimate
the most probable value for variable $F_5$ given the observed values for $F_1,F_3,F_4$, we can calculate $P(F_5|F_1,F_3,F_4)$ as: 
 \begin{eqnarray}
 &&  \hspace{0.8cm}P(F_5|F_1,F_3,F_4)   =  \frac{P(F_1,F_3,F_4,F_5)}{P(F_1,F_3,F_4)} \nonumber \\
                                      && \hspace{1.3cm}=   \frac{\displaystyle \sum_{F_2} P(F_1,F_2,F_3,F_4,F_5)}{\displaystyle \sum_{F_2,F_5}P(F_1,F_2,F_3,F_4,F_5)} \nonumber \\
                                       && =   \frac{\displaystyle \sum_{F_2} P(F_1|F_4)P(F_2)P(F_3|F_5)P(F_4)P(F_5|F_2,F_4)}{\displaystyle \sum_{F_2,F_5}P(F_1|F_4)P(F_2)P(F_3|F_5)P(F_4)P(F_5|F_2,F_4)} \nonumber \\
                                       &&=  \frac{P(F_1|F_4)P(F_3|F_5)P(F_4)\displaystyle \sum_{F_2}P(F_2)P(F_5|F_2,F_4)}{P(F_4)P(F_1|F_4)\displaystyle \sum_{F_2,F_5}P(F_3|F_5)P(F_2)P(F_5|F_2,F_4)}  \nonumber \\
 \end{eqnarray} 
 
Summing out the unobserved features and ``pushing in" the factors  in the sums is known as {\em variable elimination} \citep{Pearl:94}, which is the simplest exact inference algorithm.

\vspace{0.1cm}
In this work we use Gaussian nodes inference \citep{Shachter1989}. Gaussian nodes are commonly used for continuous data, each variable is modeled with a Gaussian distribution where its parameters are linear combination of the parameters of the parent nodes in the Bayesian network. Let $F_j$ be a node with $p$ parents, where each parent has a Gaussian distribution with mean $\mu_i $ and variance 
$\sigma_i$  ($i \in [1 \ldots p]$). We model $F_j$ with a Gaussian distribution with mean $\mu = [ \mu_1, \ldots , \mu_k ]$ and covariance matrix $ \Sigma = [ \sigma_{ik} ]$, where $\sigma_{ik}$ is the covariance between the $i$-th parent of $F_j$ and the $k$-th parent of $F_j$. The probability distribution for node $F_j$ is:
  
\begin{equation}
P(F_j) =  \mathcal{N} ( \beta_0 + \beta^T \mu ; \sigma^2 + \beta^T \Sigma \beta ),
\label{eq:LinearG}
\end{equation}
 
  Where $\beta$ and $\sigma$ are the parameters of the linear combination (which need to be estimated in the learning process). In section \ref{sec:ParLearnComplete} we explain details about Gaussian nodes representation and how to estimate the parameters. For the scope of this section, we assume that the parameters are known.\\


The simple idea behind inference with Gaussian nodes is that features that are not involved in the calculus of a probability can be eliminated from the Bayesian network (barren nodes). The easiest barren nodes to be eliminated are leaf nodes because they can be deleted without doing any other change in the network. Unfortunately, not all barren nodes are leaves. The key idea is to perform arc reversals in order to let barren nodes as leaves. When such a reversal is performed,  to preserve the joint distribution the  network has to be adjusted, or re-learned.
\cite{Shachter1989} describes a methodology for adjusting the network parameters that we adapted for this work too (see Appendix \ref{app:ArcRev} for details).

\subsection{Learning Bayesian Networks with complete data} 
 \label{sec:BNi2}
In previous section we showed how to make inference once the BN is known. 
Learning the network involves  learning the structure (edges) and  the parameters (probability distributions on each of the factors). 
We explain both cases separately in the next subsections.
 
\subsubsection{Structure Learning with complete data}\label{sec:LearnStructComplete} 
 
Given that the number of possible network structures grows exponentially with the number of nodes or features \citep{Cooper1992}, it is not possible to do an exhaustive search. 
Usually a greedy search strategy it is necessary to find a suitable solution, in this work we use the K2 algorithm \citep{Cooper1992}. Starting with an initial random order of features (nodes) and an empty network (no edges), we start adding parents to each variable, such that the next parent we add is the one who creates the highest improvement in the network score, we keep adding parents until we complete the maximum allowed (parameter given by the user). In our work we use a maximum of three parents. Note that if one node has already two parents and we attempt to add the third one, it might be possible that keeping two parents is better than adding a third one, in that case the node stays with two parents. 

The score of a network structure is related to how the structure fits data. To calculate the score, we evaluate the probability of the structure given the data, which corresponds to apply the same factorization imposed by the structure and use multinomial distributions over each factor ($P(F_{j} |  {\Pa}_{BN}(F_j)  )$ in eq.~\ref{Eq:BN_Fact}). We estimate each probability by firstly discretizing the possible values  that each feature $F_j$ can take,  $(f_{j1},\ldots,f_{jr_{j}})$  and then  creating a multi-dimensional histogram for $P(F_{j} |  {\Pa}_{BN}(F_j) )$.
  
\begin{figure}
    \centering
    \includegraphics[width=7.5cm,height=4cm]{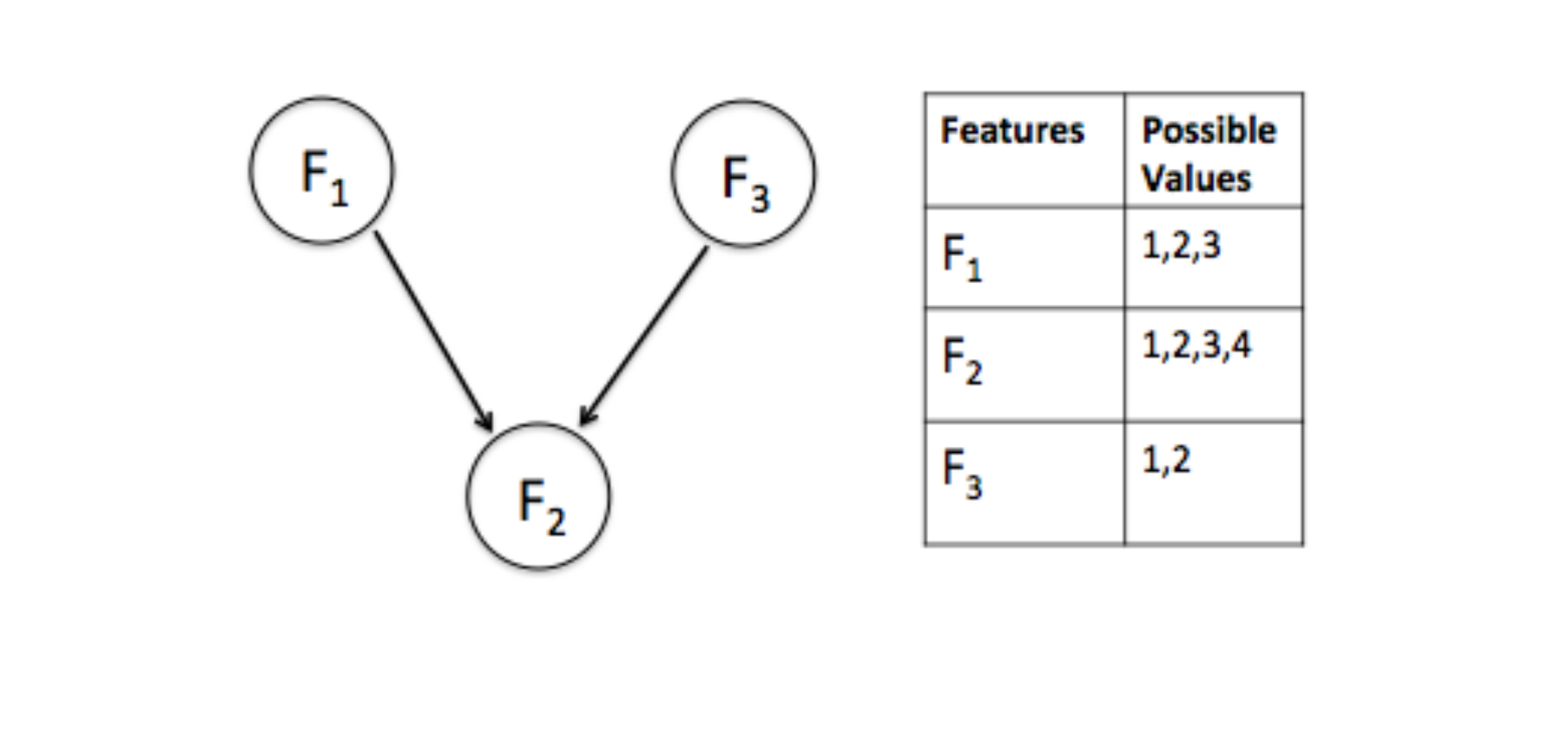}
    \caption{Example of a simple Bayesian network with features $\{ F_1, F_2 ,F_3 \}$ and the values each feature can take.}
    \label{Fig:Notation_example}
 \end{figure}

  Consider the feature $F_j$. Let $q_j$ be the number of possible instantiations of the parents set ${\Pa}_{BN}(F_j)$. Recall $r_{j}$ be the maximum number of values that variable $F_j$ can take. Let $N^j_{k,m}$ be the number of cases in data where variable $F_j$ has the value $f_{jk}$ ($k \in [1 \ldots  r_{j}]$)  when its set of parents ${\Pa}_{BN}(F_j)$  is instantiated to some value $w^j_{m}$,
  and let $N^j_{m} = \sum_{k=1}^{r_j} N^j_{k,m}$. For example, in figure \ref{Fig:Notation_example}, if $j=2$, $[ f_{j1} = 1 , f_{j2} = 2 , \ldots , f_{j4} = 4]$, $q_j = 6$ ($3 \times 2$ possible values of the joint combination of parents), $ w^j_{1} = \{ 1 \quad 1\} , w^j_{2} = \{ 1 \quad 2\} , \ldots , w^j_{6} = \{ 3 \quad 2\}$. Note that $m$ is an index moving in the possible combinations of values of the joint set of parents ($m \in [1 \ldots 6]$).\\  

Then the probability of a given structure $B_s$ can be shown to be given by  (see Appendix \ref{app:bnscore} for a derivation):\\    
 \begin{eqnarray} \label{eq:BN_Score}
   P(B_s | \data) &=& P(B_s) \prod_{j=1}^{D} \prod_{m=1}^{q_j} \frac{(r_j - 1)!}{(N_{jm} + r_j - 1)!} \times \nonumber \\
                            &&  \times \prod_{k=1}^{r_j} N_{jmk}! 
   \end{eqnarray}
 
\noindent where the term $P(B_s)$ is the prior on the network structure $B_s$.

\subsubsection{Parameter Learning with complete data} \label{sec:ParLearnComplete} 

Learning the parameters of a BN means to learn the distribution of each of the factors in the right side of equation \ref{Eq:BN_Fact}. The factorization of equation \ref{Eq:BN_Fact} is given by the structure of the BN. This tells us that to learn the parameters it is necessary first to know the structure.

   Given that features involved in our work are all continuous, we again use Gaussian nodes  \citep{Shachter1989}. 
   
   Let $F_j$ be a node in the network and ${\Pa}_{BN}(F_j)$ the set of parents for $F_j$. Lets assume that $F_j$ has $k$ parents ($| {\Pa}_{BN}(F_j) | = k$) and we model $F_j$ as a linear Gaussian of its parents:
  
\begin{equation}
P(F_j) =  \mathcal{N} ( \beta_0 + \beta^T \mu ; \sigma^2 + \beta^T \Sigma \beta ),
\label{eq:LinearG}
\end{equation}

 \noindent where the set of parents ${\Pa}_{BN}(F_j)$ are jointly Gaussian $\mathcal{N}(\mu ; \Sigma)$
and $\mu, \Sigma$ are calculated from the data. 
 Note that $\mu$ and $\beta$ are $k$ dimensional vectors, and the matrix $\Sigma$ is $k \times k$.
 
  To learn a Gaussian node, we learn the set of parameters $\{ \beta_0, \ldots , \beta_k ; \sigma \}$ of the linear combination. Let ${\Pa}_{BN}(F_j) =  \{ \tilde{F}_1, \ldots, \Fd_k \}$ be the parent nodes with respective means $\{ \mu_1, \ldots, \mu_k \}$, then  $P(F_j | {\Pa}_{BN}(F_j) ) = \mathcal{N} (\beta_0 + \beta_1 \tilde{F}_1 + \cdots + \beta_k \tilde{F}_k ;  \sigma^2)$. 
   Our task is to learn the set of parameters $\theta_{F_j} = \{ \beta_0, \ldots , \beta_k ; \sigma \}$. To learn
those parameters we optimize the log-likelihood, expressed as:

  \begin{eqnarray} 
    l_{F_j}(\theta_{F_j} | \data) &=& \sum_{i=1}^{n} \left[ -\frac{1}{2} \log(2\pi \sigma^2) - \frac{1}{2\sigma^2} (\beta_0 + \beta_1 x_{i \tilde{F}_1} +  \right. \nonumber\\
                                                 && \left. \cdots  + \beta_k x_{i \tilde{F}_k} - x_{ij})^2 \right].
  \label{Eq:GN_lik}       
  \end{eqnarray} 
   
  \noindent by setting its derivative with respect to $\beta_0$ to zero. We have:
   
  \begin{eqnarray} 
     E[F_j] &=& \beta_0 + \beta_1 E[\Fd_1] + \cdots + \beta_k E[\Fd_k]         ,         
  \label{Eq:GN_1}       
  \end{eqnarray} 
	   
   \noindent where $E[F_j]= \mu_j$ is the expectation of the variable $F_j$ in the data.
   Setting the derivative of eq.~\ref{Eq:GN_lik} with respect to  $\beta_1, \ldots , \beta_k$ to zero we have the following $k$ equations:
  \begin{eqnarray} 
     E[F_j \cdot \Fd_1 ] &=&  \beta_0 E[\Fd_1] + \beta_1 E[\Fd_1 \cdot \Fd_1]  + \cdots   \nonumber\\
                                        &&      \cdots  +  \beta_k E[\Fd_k \cdot \Fd_1]\\
       & \vdots & \nonumber\\      
     E[F_j \cdot \Fd_k ] &=&  \beta_0 E[\Fd_k] + \beta_1 E[\Fd_1 \cdot \Fd_k]  +  \cdots  \nonumber \\
                                     && +  \cdots  +  \beta_k E[\Fd_k \cdot \Fd_k]
  \label{Eq:GN_2}       
  \end{eqnarray} 

  Setting the derivatives to zero, we end with $k+1$ linear equations with $k+1$ unknowns. We can solve the equations using standard linear algebra to find the $k+1$ solutions $\beta^*_0 , \ldots , \beta^*_k$.
  To find $\sigma$ we replace the values of $\beta^*_0 , \ldots , \beta^*_k$ in eq.~\ref{Eq:GN_lik} and set the derivative of the log likelihood with respect to $\sigma^2$ to zero, then we have:  
    
    \begin{eqnarray} 
    \sigma^2 = \Cov[F_j,F_j] - \sum_{p=1}^{k} \sum_{q=1}^{k} \beta^*_p \beta^*_q \, \Cov[\Fd_p,\Fd_q]    
  \label{Eq:GN_3}       
  \end{eqnarray}     

\noindent Where $\Cov[\Fd_p , \Fd_q] = E[\Fd_p \cdot \Fd_q ] - E[\Fd_p] \, E[\Fd_q ] $.
Note that if parent nodes are root nodes, they are just modeled with a unidimensional normal distribution
 and eq. \ref{Eq:GN_1}, \ref{Eq:GN_3} return the mean and variance of that variable.

\subsection{Learning Bayesian Networks with missing data} 
 \label{sec:BNi3}
 Once we know how to learn the structure and the parameters of a Bayesian network under complete data, we now turn our attention on how to  learn both the structure and the parameters under incomplete data. To learn the parameters with missing data, we need to previously know the structure and to learn the structure we need to guess the missing values. This is done in a iterative method.

We start first describing the parameter learning algorithm and then the structure learning model.  
\subsubsection{Learning parameters with missing data} \label{sec:LearnParMissing}
  
   We assume that we already know the structure of the Bayesian network before we start learning
   the  distribution parameters with missing data. The basic idea is to start estimating the joint
   distribution of the set of (root) parent nodes  from incomplete data using a multivariate Gaussian distribution. 
   Then we estimate the distribution of children
   nodes like in the complete data case (section \ref{sec:ParLearnComplete}) sampling the missing
   values of the parents from the distributions learned at the beginning. 
   
   To learn the parameters of a multivariate Gaussian distribution in the incomplete data case, 
   we use the method proposed in \citet{Ghahramani:1995}. 
   To optimize the log likelihood of the model given the data under the missing data case,   
   each data point $x_i$ can be written as
   $x_i = \{ x_{i}^{o},x_{i}^{m} \}$, using the super scripts $m$ and $o$ to indicate the features that are observed or missing. 
    Let $\Sigma = \{ \Sigma^{oo} , \Sigma^{om} , \Sigma^{mo}, \Sigma^{mm} \}$ and $\mu = \{ \mu^m , \mu^o  \}$ be the covariance 
    matrix and vector mean of the multivariate Gaussian distribution we are estimating.   
   The log likelihood for incomplete data, including the new notation for $x_i$ and using  a mixture of Gaussians distribution can be written as:
   
 \begin{eqnarray} 
     l(\theta | x^{o},x^{m}) & = & \sum_{i=1}^{n} [ \frac{n}{2} \log 2 \pi + \frac{1}{2} \log | \Sigma| \nonumber\\
                                           &   &  -\frac{1}{2} (x_{i}^{o} - \mu^{o})^T  \Sigma^{-1,oo}  (x_{i}^{o} - \mu^{o}) \nonumber\\ 
                                           &   &  - (x_{i}^{o} - \mu^{o})^T  \Sigma^{-1,om}  (x_{i}^{m} - \mu^{m}) \nonumber\\
                                           &   &  -\frac{1}{2} (x_{i}^{m} - \mu^{m})^T  \Sigma^{-1,mm}  (x_{i}^{m} - \mu^{m})] \nonumber\\
  \label{Miss:lik}       
 \end{eqnarray}

 Given that we have the likelihood expressed in terms of unknown latent features (the unobserved part of data), we optimise it using the expectation maximization (EM) algorithm \citep{Dempster:1977}.  EM optimizes the likelihood function of a model which depends on latent or unobserved features. The optimization procedure is a two step iteration. First step, the expectation step (E-step), calculates the expected value of the latent features to be used in the likelihood function. 
    In other words,  the E-step creates a function to be optimized, using the expected value of the latent features estimated from the current value of the unknown parameters. The maximization step (M-step) is the maximization of the likelihood function created by the E-step, generating a new value of the current parameters (to be used again in E-step).    

   In the E-step we need to estimate the unobserved part $x_{i}^{m}$. We can express the expected value of $x_{i}^{m}$ as:
   
 \begin{eqnarray}    
     E[x_{i}^{m} | x_{i}^{o} , \mu,\Sigma] &=& \mu^{m} +  \Sigma^{mo} \Sigma^{-1,oo} (x_{i}^{o} - \mu^{o})
     \label{Eq:Missing_Expec}
 \end{eqnarray}\\
   
  Starting for an initial guess of the parameters $\mu$ and $\Sigma$, we calculate the expected value of missing data using equation \ref{Eq:Missing_Expec}, then we optimise the values of $\mu$ and $\Sigma$ and we continue iterating until $\mu$ and $\Sigma$ do not change substantially.
  
   After estimating the joint Gaussian distribution of the set of parents, we can estimate the Normal distribution of the children like in the complete data case (sec. \ref{sec:ParLearnComplete})
   where the missing values of the parent are sampled from the learned multivariate Gaussian.

   
 
   

\subsubsection{Learning the network structure with missing data} 

  To learn the structure of a Bayesian network with missing data, we complete the missing values and then iterate to improve these values using the structure learned so far \citep{Singh1997}.   
  Algorithm \ref{alg:Learn_BN_Missing} shows the main steps to construct a Bayesian network structure from missing data.\\ 

\begin{algorithm}
\begin{minipage}[t]{0,48\textwidth}
 \label{alg:Learn_BN_Missing}
 \caption{Algorithm to Learn BN structure with missing data}
    \emph{$\bullet$ Learn for each variable in $\{ F_1, \ldots , F_D \}$ an univariate Gaussian Mixture $GM(i) \; i \in [1 \ldots D]$}\;
     \BlankLine
     \emph{$\bullet$ Create $M$ complete datasets $D_{s}^{1}, \; s \in [1 \ldots M]$ filling the missing values of each variable $F_i$ with values sampled from $GM(i)$}\;
     \BlankLine
     $\bullet \quad t = 1$\;   
     \BlankLine
    \While{ Convergence criteria is not achieved }
     {
         \For{ $s = 1$ to $M$}
          {
             From each complete dataset in $D_{s}^{(t)}$, learn a BN, $B_s^{(t)}$ (sec. \ref{sec:LearnStructComplete}).       
          }
      \emph{$\bullet$ Create one Bayesian network structure $B^{(t)}$ as the union of all the BNs} \footnote{The union is performed in two steps: i) make all structures $B_s^{(t)}$ consistent with the features order used in the algorithm from sec. \ref{sec:LearnStructComplete} by performing arc reversals and ii) create the arc union of all the consistent structure from the previous step}  \;
      \BlankLine
      \emph{$\bullet$ Learn the parameters $\theta^{(t)}$ using the original incomplete data and the network structure $B^{(t)}$ (section \ref{sec:LearnParMissing})}\;
      \BlankLine
      \emph{$\bullet$ Use the network $\langle B^{(t)},\theta^{(t)} \rangle$ to sample new values and create new completed datasets $D_{s}^{(t+1)}$}\;
      \BlankLine
     $\bullet \quad t = t + 1$\;
     }
\end{minipage}     
\end{algorithm}

 The convergence criteria is that the score of the network $\langle B^{(t)},\theta^{(t)} \rangle$ does not change substantially. Note that in the first iteration we just fill the missing values using an independent Gaussian mixtures model. Although independency is a very strong assumption, in our case does not affect the final result given that in all subsequent steps we re-fill the missing values with data sampled from the current Bayesian Network, in other words we use all probabilistic dependencies between features given by the network structure. 
   
\section{The automatic classification model}
  \label{sec:class}
  
In previous sections we show how to fill missing values using probabilistic dependencies between features. After we infer the missing values using the Bayesian network, we proceed to train the automatic classifier using the new training completed set. In this work we use a Random Forest (RF) classifier \citep{Breiman2001}, which is a popular and very efficient algorithm based on decision tree models \citep{Quinlan:1993} and Bagging for classification problems \citep{Breiman:1996, Breiman:2001} \footnote{Other models can be used for classification, but we found that RF gives superior results}.
 It belongs to the family of ensemble methods, appearing in machine learning literature at the end of nineties \citep{Dietterich:2000} and has been used recently in the astronomical journals \citep{Pichara:2012,Carliles:2010,Richards:2011}. We give a very brief explanation here of how RF works; the reader can find detailed description in \cite{Breiman2001}.

The process of training or building a RF given training data
is as follows:

\begin{itemize}
  \item Let $P$ be the number of trees in the forest (model parameter)
  and $F$ be the number of features describing data.
  \item Build $P$ sets of $n$ samples taken with replacement from the training set; this is called bagging. Note that each of the $P$ bags has the same number of elements with the training set but less different examples, given that the samples are taken with replacement (The training set also has $n$ samples).

   \item For each of the $P$ sets, train a decision tree (without prunning) using at each node a random sample of $F' \leq \leq F$ possible features to select the one that optimises the split. ($F'$ is a model parameter)
\end{itemize}

  The RF classifier creates many linear separators, attempting to separate between elements of different classes using some features (the ones given by the nodes of each decision tree) and some data points (the ones given by each of the bags).

  
     Each of the decision trees creates one decision and the final decision is the most voted class among the set of $P$ decision trees (see \citet{Breiman:2001} for more details).  
 Worth noting,  \cite{Breiman:2001}  showed that as the number of trees goes to infinity the classification error of the RF becomes bounded and the classifier does not overfit the data.

\section{Experimental Results} \label{sec:Experiments}

  In this section we show the results from the application of the model on four astronomical catalogs, three with missing values. We not only show the advantages of the model, but we produce a 
catalog of variable stars within the LMC which is available for downloading\footnote{http://iic.seas.harvard.edu/research/time-series-center}.

  First, we prove the imputation accuracy of our model performing imputation tests in real datasets. Then, to test the advantage of the model, we proved three main facts: i) it is possible to learn an automatic classification model that is able to deal with missing data, ii) information with missing data can be useful for automatic classification, in other words the model outperforms any model that uses a subset of training set with complete data, iii) the proposed model overcomes the case where missing data is filled using traditional statistical methods that model each variable independently. 
   
   Fortunately the main computational cost of the algorithm occurs during the training phase, where the model needs to learn the Bayesian network structure and parameters. After training the model, to perform the inference on missing values for a lightcurve takes a fraction of a second. 
   
   We use three astronomical catalogs with missing data, SAGE \citep{Meixner2006},  UBVI \citep{Piatti2011} and 2MASS \citep{2MASS:2006}. We also use the MACHO catalog \citep{Alcock1997ApJa}, with no missing values, but  useful in comparing the lightcurve classification accuracy between the MACHO features  with the additional incomplete extra features from SAGE, UBVI and 2MASS. We process around 20 million of objects. The MACHO lightcurves are described using 14 variability features:  CAR $\sigma$, Mean Mag, CAR $\tau$, $\sigma$, $\eta$, Con, Stetson L,  CuSum, B-R, Period, Period SNR, Stetson K AC, N above 4, N below 4. See \citep{Pichara:2012} for a description of the MACHO features.

\subsection{Imputation Tests}
 
  To calculate the imputation error, we use the MACHO dataset where we randomly delete  5\%, 10\%, 15\% and 20\% of data entries in order to simulate missing values. We run our model and  
  measure the $NRMSE$ (Normalized Root Mean Squared Error) over the predicted values, defined as:
  
  $$NRMSE = \sqrt{ \frac{Mean([x_{imp} - x_{true}]^2)}{var(x_{true})} }$$

  Where $x_{imp}$ is the imputed value and $x_{true}$ is the true value. When the estimation is accurate, $NRMSE$ approaches  $0.0$  where when the estimation is equivalent to a random guess, NRMSE approaches to 1.0.  
  We compare our imputation results with an imputation method using mixtures of Gaussians. Table \ref{tab:Imp}  shows our results.

\begin{table} [h!] 
\begin{center} 
\caption{NRMSE using Bayesian Networks and Gaussian Mixtures in MACHO dataset. Missing values were artificially generated completely at random}
\label{tab:Imp}
\begin{tabular}{|ccccc|}
\hline
                   & \textbf{5\%} & \textbf{10\%} & \textbf{15\%} & \textbf{20\%}\\
\hline                   
BN             & 0.396 & 0.437 &  0.514  & 0.523\\
Gaussian & 0.624 & 0.701 &  0.764 &  0.790\\ 
\hline
\end{tabular}  
\end{center}  
\end{table}

  We can see from table \ref{tab:Imp} that BN method present less NRMSE compared with Gaussian Mixtures Imputation.

 \subsection{Interpreting the BN}
 
\begin{figure*}
    \centering
    \includegraphics[width=13cm,height=10cm]{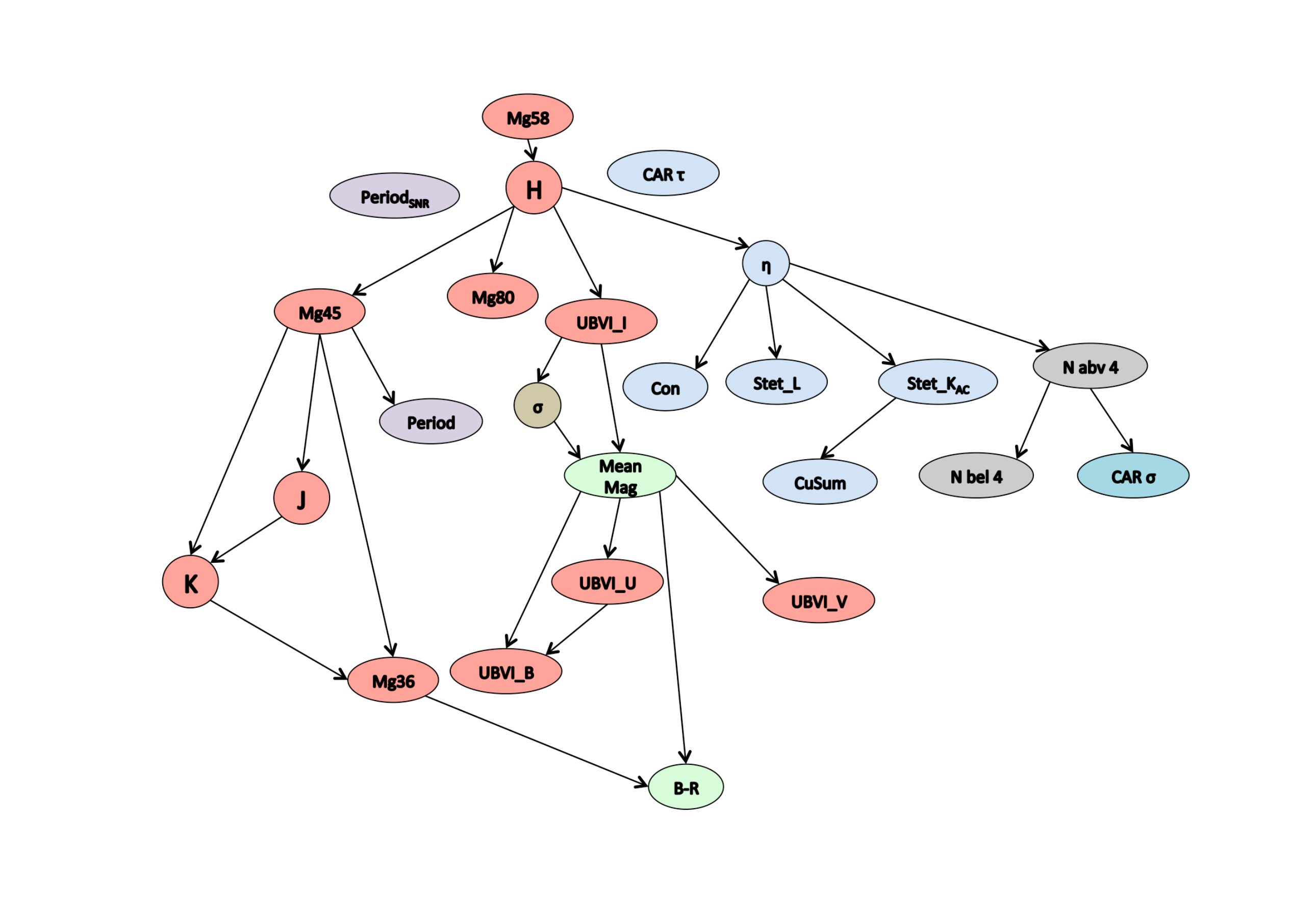}
    \caption{Bayesian Network structure for the MACHO dataset, colours indicate that the nodes belong to the same type of features.}
    \label{Fig:BN_MACHO}
 \end{figure*}

   One of the advantages of Bayesian networks is that they provide a conditional probability structure of features, which can be interpreted to attain deeper insight into your data. Figure   
    \ref{Fig:BN_MACHO} shows the BN structure our model found for the MACHO dataset. Connection among features indicates a degree of probabilistic dependency among features. Nodes
    that are not connected with any other nodes are estimated independently from the others. We added colours to the nodes to indicate groups of features that belong to the same \lq\lq type" of features.
     For example, features related to the magnitude level of the object are in red, as we can see, there are many connections among these kind of nodes, showing that the learning algorithm 
     despite the missing data was able to detect most of the dependency relationships. There are some relationships that the model could not find, for example, the feature CAR $\tau$ was modelled
     as independent given that is not connected with any other feature in the network structure.

 \subsection{Classification results in missing data catalogs}
 
  For SAGE and 2MASS we used a training set of 1955 objects described in 7 features ($J, H, K, \rm{m}_{36}, m_{45}, m_{58}, m_{80}$). Table~\ref{Table:SAGE} shows the percentage of missing values for different features in SAGE/2MASS catalogs.
  
  \begin{table}
 \caption{Percentage of missing values on SAGE/2MASS catalogs}  
  \begin{center}
   \begin{tabular}{|cc|}
   \hline
    Vars  & \% of missing values\\ 
   \hline  
     $J$ & 54\%\\
     $H$ & 54\% \\
     $K$ & 58\%\\
     $ m_{36}$ & 1\%\\
     $ m_{45}$ & 13\%\\
     $ m_{58}$ & 68\%\\
     $ m_{80}$ & 74\%\\
     \hline
  \end{tabular}
      
  \end{center}
  \label{Table:SAGE}      
  \end{table}
  
   For UBVI catalog \citep{Piatti2011} we used 4193 training instances described in 4 features ($U,V,B,I$). Table \ref{Table:UBVI} shows the percentage of missing values for different features in UBVI training set.

   \begin{table}  
   \begin{center}
   \caption{Percentage of missing values on UBVI catalog} 
    \begin{tabular}{|cc|}
    \hline
     Vars  & \% of missing values\\ 
      \hline
     $U$ & 49\%\\
     $B$ & 0\% \\
     $V$ & 0\%\\
     $I$ & 14\%\\
     \hline
  \end{tabular}
          
    \label{Table:UBVI}   
    \end{center}
  \end{table}

  We created one training set gathering all SAGE, 2MASS and UBVI training sets. The resulting training set contains 
  seven classes of stars: Non-Variables, Quasars (QSO), Be Stars, Cepheids, RR Lyrae, Eclipsing Binaries (EB) and Long Periodic Variables (LPV). The number of objects per class on SAGE-2MASS-UBVI training set is described in table \ref{Table:Classes}.

  \begin{table}
  \begin{center}
  \caption{Number of objects per class on SAGE-UBVI training set}    
    \begin{tabular}{|cc|}
    \hline
     Class  & num. of training objects\\ 
     \hline
     Non-Variables & 1136\\
     QSO & 45\\
     Be star & 76\\
     Cepheid & 70\\
     RR Lyrae & 69\\
     EB & 100\\
     LPV & 337\\
     \hline
   \end{tabular}    
    \label{Table:Classes}     
    \end{center}  
  \end{table}  

 To evaluate the capabilities of our model dealing with missing data, we compared the results of classification accuracy of our model versus filling the missing data with independent Gaussian Mixtures model 
 on each variable. Then we take samples from that distribution to replace the missing values. 
 
 These experiments allows us to show the importance of analyze the dependency relationship between features in order to make inference on missing values. To measure the accuracy we use precision, recall and F-Score, defined as:\\
 
   F-Score $ =  2 \times \frac{{\rm precision} \times {\rm recall}}{{\rm precision + recall }},$\\   
   
\noindent where precision and recall are defined as:\\
   
    ${\rm precision} = {\rm \frac{TP}{TP + FP} }   \qquad   {\rm recall} = {\rm \frac{TP}{TP + FN}},$\\  
  
\noindent where TP, FP and FN are the number of true positives, false positives and false negatives respectively. Note that all these values are obtained using a 10-fold cross validation process.\\

  Table \ref{Table:Accuracy-Gaussian-Fill} shows the accuracy for the model which uses Gaussian Mixtures and the accuracy of the proposed model. We can see that our model
  present better results in most of the classes (bold numbers), for example, we can see significant improvement in the recall of quasars, which indicates that the model is able to detect more quasars that the model
  which fills features independently. We also increase the precision and recall for Be stars, RR-Lyraes, and Long Periodic Variables.


\begin{table*}
  \caption{Precision, Recall and F-Score for different classes using two different methods for filling missing values in SAGE-2MASS-UBVI training set. 1) Independent Mixtures of Gaussian and 2) Our model.} 
   \begin{tabular}{|c|ccc|ccc|}
   \hline   
   & & Gaussian mixture & &  & Our model &\\
  Class &   Precision  & Recall  & F-Measure  &      Precision  & Recall  & F-Measure  \\
       \hline	
       None Variables & 0.857   &  0.952  &   0.902 &  \textbf{0.878}   &  0.942  &   \textbf{0.909}    \\
       Quasar & 0.9     &  0.8   &         0.847 &   0.878  &  \textbf{0.956}   &   \textbf{0.915}\\
       Be & 0.679  &   0.5 &        0.576   & \textbf{0.724}  &   \textbf{0.553} &    \textbf{0.627} \\
       Cepheid & 0.805   &  0.886 &   0.844   & 0.785   &  0.886 &    0.832   \\
       RR-Lyrae & 0.333   &  0.014 &   0.028  &  \textbf{0.583}   &  \textbf{0.203} &    \textbf{0.301} \\
       EB & 0.5    &   0.25  &       0.333  & \textbf{0.525}    &   \textbf{0.31}  &   \textbf{0.39} \\
       LPV & 0.919  &   0.938  &   0.928  & \textbf{0.925}  &   \textbf{0.947}  &   \textbf{0.935}  \\
       \hline
       Weighted Average:  & 0.821 & 0.851 & 0.826 & \textbf{0.846} & \textbf{0.863} & \textbf{0.848}\\
      \hline   
   \end{tabular}    
       \label{Table:Accuracy-Gaussian-Fill}    
  \end{table*}


  After evaluating our proposed method with SAGE, 2MASS and UBVI catalogs, we aim to probe that all the information encoded by these catalogs is useful to classify variable stars after missing data were imputed. 
   To probe that, we combined again the SAGE-2MASS-UBVI training set with a training set used by \citet{Pichara:2012} in a previous work, the MACHO catalog \citep{Alcock1997ApJa}. In the previous work, an automatic classifier was built in order to detect quasars in MACHO database. This training set was created extracting 14 time series features per band (see \citet{Pichara:2012} for further details). To evaluate the contribution of our model we trained a new classifier which uses the previous 14 features from \cite{Pichara:2012} and the new features from SAGE-2MASS-UBVI training set, after our model processed the missing values. We expect that the new classifier improves the quasar classification showing higher recall and precision values in the training set and getting a new high quality list of quasar candidates. Table \ref{Table:MACHO_SAGE_UBVI_Train} shows the results of both training set, with and without SAGE-2MASS-UBVI catalog, showing that we improve the value of F-Score in quasar detection. 
   
\begin{table*}
  \caption{Precision, Recall and F-Score to compare the model used \citet{Pichara:2012} with and without the information of SAGE-UBVI training set}
   \begin{tabular}{|ccc|}
  \hline   
      &  Adding SAGE-2MASS-UBVI features & Only MACHO Features\\
       \hline	
       Quasar Precision   &  0.843 & 0.857\\      
       Quasar Recall  &   \textbf{0.956} & 0.8\\
     Quasar F-Score  &  \textbf{0.896} & 0.828\\
   \hline   
   \end{tabular}    
       \label{Table:MACHO_SAGE_UBVI_Train}    
  \end{table*}

 Moreover, after training the model, we run it on the whole MACHO catalog, in order to generate a new quasar candidate list. To evaluate the quality of the new quasar candidate list, we calculate the matching level of our list of candidates with the previous known lists. We use the recent works \citep{Kim:2012,Pichara:2012} to compare their candidates with the list proposed in this work. \cite{Kim:2012} found a list of 2566 candidates and a refined list of 663 strong candidates. In \citet{Pichara:2012}, we found a list of 2551 candidates, with 74\% of matches with the 663 refined strong candidates. In this work we improve the list, getting a list of 1730 candidates, from where we got 562 matches with the previous list of 2566 candidates and 502 matches with the previous list of 663 strong candidates (75.7\%). We can see that our list of 1730 candidates has about the same level of matching with the previous strong candidate list but reducing the size of the list by a 32\%.
  

\section{Conclusions}
\label{sec:con}
We show a new way of dealing with missing data, testing on real astronomical datasets, showing that catalogs with miss- ing data can be useful for automatic classification. One of the main advantages of our model is that it makes possible to integrate catalogs in order to increase the available information for the training process. We improve the accuracy of our results in previous work on quasar detection due to the integration of new catalogs with missing data. Our model considers probability dependencies between features that make possible to take advantage of the observed values, in order to increase the accuracy of the estimation when the number of observed values increase. Most of the computational time required is during the training time that makes possible to run the model in complete catalogs because the model just need to perform inference on the missing values, which takes less than a second per object.


\section*{Acknowledgments}
This work is supported by Vicerrector\'ia de Investigaci\'on (VRI) from Pontificia Universidad Cat\'olica de Chile, Institute of Applied Computer Science at Harvard University, and the Chilean Ministry for the Economy, Development, and Tourism's Programa Iniciativa Cient\'{i}fica Milenio through grant P07-021-F, awarded to The Milky Way Millennium Nucleus.

\bibliography{../Bibtex/library}
\appendix
\section{Arc Reversal}
\label{app:ArcRev}
\begin{figure}
  \begin{center}
    \centering
    \includegraphics[width=9cm,height=4.5cm]{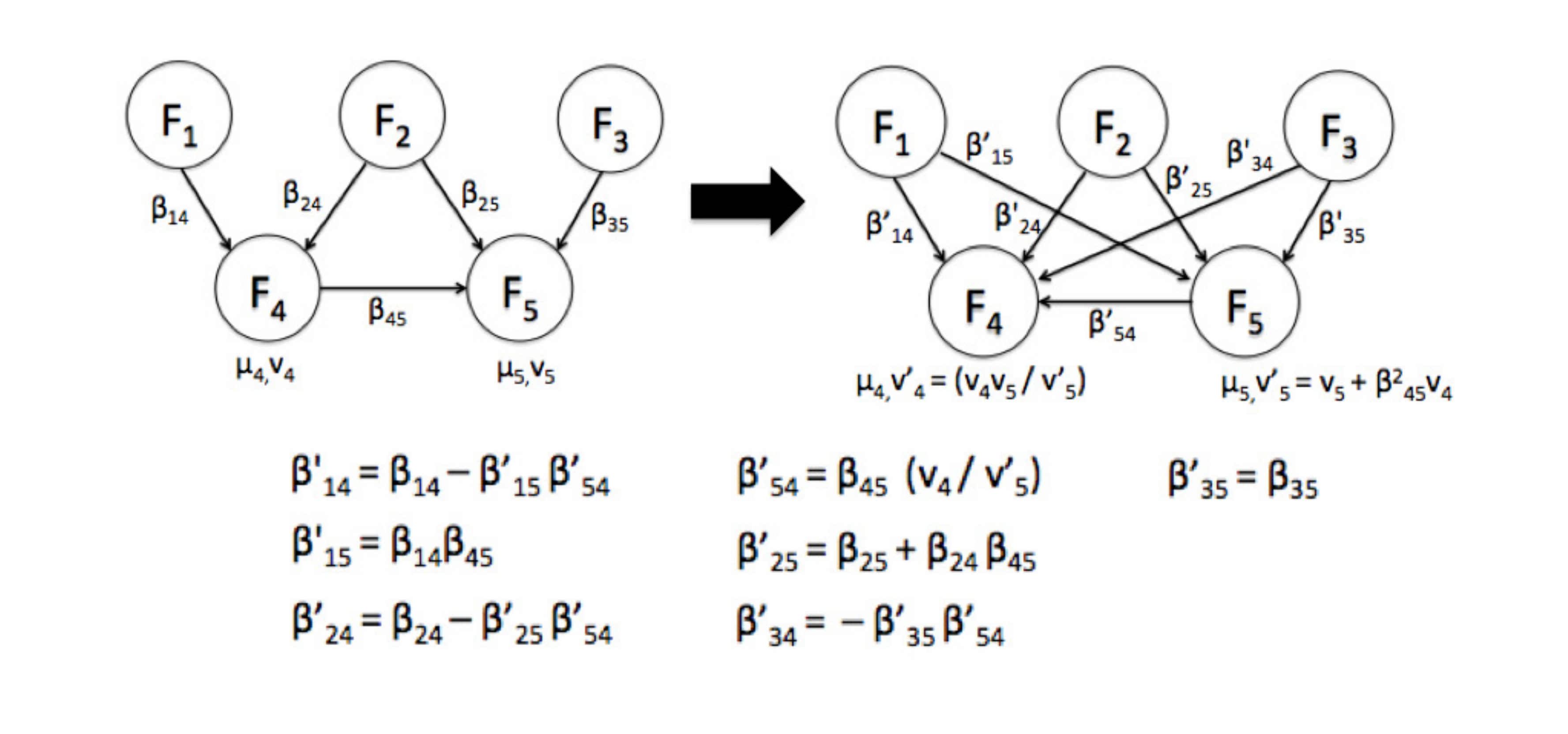}
    \caption{Arc reversal example for Gaussian nodes. $\beta$ values correspond to the coefficients of the linear combination.}
    \label{Fig:Arc_Reversal}
  \end{center}
 \end{figure}

To understand the meaning of arc reversal, it helps to think as each node propagates its variance downstream to its successors
(these variances are the elements of the covariance matrix in eq. \ref{eq:LinearG} as we describe in sec. \ref{sec:ParLearnComplete}).
Suppose we reverse the arc $F_i \rightarrow F_j$. Before reversing, part of the variance in $F_j$ was explained by $F_i$, then after the reversal we have to compensate this 
by adding an arc from the parents of $F_i$ to $F_j$.
Also, part of the variance of $F_i$ is now explained by $F_j$, so $F_i$'s new variance must be discounted in order to adjust for that value. Figure \ref{Fig:Arc_Reversal} shows an example of the arc reversal procedure and updates of variances. In that example $F_4$ is a barren node but can not be remove from the network for it is not a leaf node.
By reversing the arc $\{F_4\rightarrow F_5\}$ to  $\{F_5\rightarrow F_4\}$ the variances and edge parameters are
adjusted as described in the figure. With the arc reversed $F_4$ is now a leaf node and can be removed from the network. 

 Formally, suppose we want to perform inference to calculate $P(F_J | F_K)$, where $F_J$ and $F_K$ are sets of features such that $F_J \cap F_K = \emptyset$. Let $N$ be the total set of nodes in the network ($(F_J \cup F_K) \subset N$). We first create an ordered sequence $s$ of nodes in $N$ such that $F_K \prec_{s} F_J  \prec_{s} N \setminus (F_J \cup F_K)$ (Note that \lq\lq$\setminus$'' is the sets subtraction operator). We use the notation $\prec_{s}$ just to define an order relationship between the elements inside the sequence $s$, then if $F_K \prec_{s} F_J$ means that $F_K$ is before $F_J$ in $s$. The idea of this order is just leave evidence nodes before in the sequence, that ensures that they will be ancestors in the network, making easier the flow of information in the graph. The steps to perform inference are:

 \begin{algorithm}
   Given the ordered sequence $\prec_{s}$:
 \For{each arc $F_i \rightarrow F_j$ in the BN}
    { 
      if $F_j \prec_{s} F_i$ then reverse the arc $F_i \rightarrow F_j$ \;
    } 
 Delete all nodes in $N \setminus (F_J \cup F_K)$ from the resulting network \;
  $P(F_J | F_K) = \frac{P(F_J , F_K)}{P(F_K)}$ \;
 \vspace{.3 cm}
$\qquad \qquad \qquad = \displaystyle \frac{  \prod_{j=1}^{N \setminus (F_J \cup F_K)}P(F_j | {\Pa}_{BN}(F_j) )}{\sum_{N \setminus F_K} \prod_{j=1}^{N \setminus (F_J \cup F_K)} P(F_j | {\Pa}_{BN}(F_j) )}$ 
\end{algorithm}

\vspace{.7cm}  The last step of the algorithm simply uses the Bayesian network factorization to calculate the desired probability $P(F_J | F_K)$. Given that the barren nodes are not longer available in the BN, the joint probability is expressed only through features in $F_J \cup F_K$. Note that nodes in $F_J$ are descendants of nodes in $F_K$, and nodes in $F_K$ are all observed, then we just instantiate them to their values and calculate directly the probability of nodes in $F_J$ given the values in $F_K$.

\section{BN Score}
In this section we show a derivation of the equation \ref{eq:BN_Score} for BN score \citep{Cooper1992}.
The score of a network structure is related to how the structure fits data. To calculate the score, we evaluate the probability of the structure given the data, which corresponds to apply the same factorization imposed by the structure and use multinomial distributions over each factor ($P(F_{j} |  {\Pa}_{BN}(F_j)  )$ in eq.~\ref{Eq:BN_Fact}). We estimate each probability by firstly discretizing the possible values  that each feature $F_j$ can take,  $(f_{j1},\ldots,f_{jr_{j}})$  and then  creating a multi-dimensional histogram for $P(F_{j} |  {\Pa}_{BN}(F_j) )$.
  
  Consider the feature $F_j$. Let $q_j$ be the number of possible instantiations of the parents set ${\Pa}_{BN}(F_j)$. Recall $r_{j}$ be the maximum number of values that variable $F_j$ can take.
  
  Let $N^j_{k,m}$ be the number of cases in data where variable $F_j$ has the value $f_{jk}$ ($k \in [1 \ldots  r_{j}]$)  when its set of parents ${\Pa}_{BN}(F_j)$  is instantiated to some value $w^j_{m}$,
  and let $N^j_{m} = \sum_{k=1}^{r_j} N^j_{k,m}$.\\

  To decide for the best structure, we need an expression for the probability of a given structure under presence of data ($P(B_s , \data)$). Given that per each
  structure we can have a different set of parameters ($\theta_s$), we need to condition in the parameters and integrate them out.

   Using multinomial
  distributions we have:\\
  
   \begin{eqnarray} 
   \label{eq:Prob_Struct_1}
   P(B_s , \data) &=& \int_{\theta_s} P(\data | B_s, \theta_s) P(\theta_s|B_s) P(B_s) d\theta_s \nonumber \\
                            &=& P(B_s) \int_{\theta_s} [\prod_{j=1}^{D} \prod_{m=1}^{q_j}\prod_{k=1}^{r_j} \theta^{N^j_{k,m}}_{jkm}  ]  \times \nonumber \\
                            & &    \times P(\theta_s|B_s)  d\theta_s \nonumber \\
                            &=& P(B_s) \int \underset{\theta_{j k  m}}{\cdots} \int \nonumber \\
                            & &   [\prod_{j=1}^{D} \prod_{m=1}^{q_j}\prod_{k=1}^{r_j} \theta^{N^j_{k,m}}_{jkm}  ]  \times \nonumber \\
                            & &   \times [\prod_{j=1}^{D} \prod_{m=1}^{q_j} P(\theta_{j1m}, \ldots , \theta_{j r_j  m}) ] \nonumber \\ 
                            & &  d\theta_{111}, \ldots , d\theta_{j k  m}  , \ldots , d\theta_{D r_j  q_j}  
   \end{eqnarray}  
  
  \vspace{.3cm}
  Assuming a uniform distribution for  $P(\theta_{j1m}, \ldots , \theta_{j r_j  m})$ we have that  $P(\theta_{j1m}, \ldots , \theta_{j r_j  m}) = C_{jm}$ (for some constant $C_{jm}$). 
  Given that $C_{jm}$ is also a density function:
  
   \begin{eqnarray} 
   \label{eq:int_density}
   \int \underset{\theta_{j k  m}}{\cdots} \int C_{jm} \; d\theta_{111}, \ldots , d\theta_{D r_j  q_j} &=& 1
   \end{eqnarray}  
  
   Solving equation \ref{eq:int_density} yields $C_{jm} = (r_j-1)!$ (see appendix of \citep{Cooper1992}). Substituting this result and using independence of terms in equation \ref{eq:Prob_Struct_1} we have that:
  
   \begin{eqnarray} 
   \label{eq:Prob_Struct_2}
   P(B_s , \data) &=& P(B_s) \prod_{j=1}^{D} \prod_{m=1}^{q_j} \int \underset{\theta_{j k  m}}{\cdots} \int \nonumber \\
                            & &  [ \prod_{k=1}^{r_j} \theta^{N^j_{k,m}}_{jkm} ] (r_j - 1)!   \nonumber \\
                            & &  d\theta_{111}, \ldots , d\theta_{j k  m}  , \ldots , d\theta_{D r_j  q_j}                            
   \end{eqnarray}  
  
  The multiple (Dirichlet) integral in equation \ref{eq:Prob_Struct_2} has the following solution \citep{Wilks:1962}:

   \begin{eqnarray} 
   \label{eq:Prob_Struct_3}
   \int \underset{\theta_{j k  m}}{\cdots} \int \prod_{k=1}^{r_j} \theta^{N^j_{k,m}}_{jkm} d\theta_{111}, \ldots , d\theta_{D r_j  q_j}  = \nonumber \\
                             \frac{\prod_{k=1}^{r_j}   N^j_{k,m}! }{(N^j_{m} + r_j - 1)!} 
   \end{eqnarray}  

  Substituting the result of equation \ref{eq:Prob_Struct_3} in equation \ref{eq:Prob_Struct_1} we have that:
 
   \begin{equation} 
   P(B_s | \data) = P(B_s) \prod_{j=1}^{D} \prod_{m=1}^{q_j} \frac{(r_j - 1)!}{(N^j_{m} + r_j - 1)!} \prod_{k=1}^{r_j} N^j_{k,m}!
   \end{equation}
  \vspace{.3cm}
 
where the term $P(B_s)$ is the prior on the network structure $B_s$. In this work we assume that all possible network structures are equally likely, so we use the same prior for all them. 
The expression $P(B_s | \data)$ is the probability of the network structure given data, in other words, how good is the fit of the network structure with data. Better the structure fits the  data, the higher the score $P(B_s | \data)$. 
Then using the previous mentioned greedy search method, we select the structure that presents higher probability among the searched ones.

\label{app:bnscore}

\end{document}